\documentclass[latex]{revtex4}%
\usepackage{amsmath}
\usepackage{graphicx}%
\usepackage{amsfonts}%
\usepackage{amssymb}
\begin{document}

\title{Angular momentum and the geometrical gauge of localized photon states}
\author{Margaret Hawton}
\affiliation{Department of Physics, Lakehead University, Thunder Bay, ON, Canada, P7B 5E1}
\author{William E. Baylis}
\affiliation{Department of Physics, University of Windsor, Windsor, ON, Canada, N9B 3P4}

\begin{abstract}
Localized photon states have non-zero angular momentum that varies with the
non-unique choice of a transverse basis and is changed by gauge
transformations of the geometric vector potential $\mathbf{a}$. The position
operator must depend on the choice of gauge, but a complete gauge
transformation of a physically distinct state has no observable effects.\ The
potential $\mathbf{a}$ has a Dirac string singularity that is related to an
optical vortex of the electric field.

\end{abstract}
\maketitle

The orbital angular momentum of optical beams has recently received
considerable attention \cite{Allen92,PhysicsToday}. \ The spin and orbital
angular momentum of a photon, as traditionally defined, cannot be cleanly
separated, although states can be constructed that have well defined total
angular momentum per photon in any specified direction. \ Without loss of
generality we choose this to be the $z$-direction, and describe the total,
spin and orbital angular momentum along this direction with quantum numbers
$j_{z},$ $s_{z}$ and $l_{z}$ respectively. \ For example, polarized paraxial
Bessel beams for which $j_{z}=0$ are superpositions of two beams: one with
$l_{z}=1$ and $s_{z}=-1,$ and one with $l_{z}=-1$ and $s_{z}=1$
\cite{Dholakia}. \ In this letter we discuss the angular momentum of photons
localized in all three spatial dimensions. \ We show that while geometric
gauge transformations can change the angular momenta of the basis states and
rotate and rescale associated singularities and optical vortices, these
features can never be completely eliminated.

In spite of an extensive literature on nonlocalizability of photons, photon
states with arbitrarily fast asymptotic power-law \cite{APS} or exponential
\cite{BBlocalization} falloff of energy density have recently been
constructed. What has not been analyzed in the past is the deviation of these
localized states from spherical symmetry and their consequent angular momentum
content.\ It has been argued that a converging or diverging one-photon state
can never be localized exactly because of mathematical limitations imposed by
quantum field theory and, for example, the Paley-Wiener theorem
\cite{BBlocalization}. \ However, a momentum-space basis of exactly localized
states such as%
\begin{equation}
\underline{\Psi}_{\mathbf{r}^{\prime},\lambda}\left(  \mathbf{p}\right)
=Np^{\alpha}e^{-i\mathbf{r}^{\prime}\cdot\mathbf{p}}\underline{e}%
_{\mathbf{p}\lambda}~. \label{localstates}%
\end{equation}
can be constructed \cite{HawtonBaylis} with contributions from all
$\mathbf{p}$, describing a photon that may be incoming or outgoing relative to
the spacetime point of localization. This orthogonal basis, while probably not
realizable as physical photon states,\ is convenient for calculation of the
probability amplitude for photon position and, as we shall show here, for the
specification of transverse bases in general.

The basis states are eigenstates of a photon position operator with commuting
components. In spite of a long history arguing against the existence of such
an operator, we found not one, but a whole family of such position operators
related by geometric gauge\ transformations, with the gauge potential defining
the rotation of the\ transverse basis about $\mathbf{p}$. \ Details, together
with explanations where arguments against their existence fail, are given in
Ref. \cite{HawtonBaylis}. We show here that the gauge choice determines the
angular momenta of the basis.

Massless particles possess only two spin states, which can be taken as
eigenstates\ of the helicity operator \underline{$\mathbf{S}$}$\mathbf{\cdot
\hat{p}.}$ The resulting coupling of spin and momentum means that the position
operator, which generates translations in momentum space, generally does not
commute with \underline{$\mathbf{S}$}. For particles of spin 1, the components
of \underline{$\mathbf{S}$} are $3\times3$ matrices that generate rotations of
the field vectors, and the position operator is therefore not simply
$i\hbar\nabla,$ where $\nabla$ is the gradient operator in $\mathbf{p}$-space,
but rather a $3\times3$ matrix. One such operator is the Pryce photon position
operator \cite{Pryce,HawtonBaylis}, known since 1948,%
\begin{equation}
\underline{\mathbf{r}}_{P}=\hbar\left(  i\underline{I}p^{\alpha}\nabla
p^{-\alpha}+\frac{1}{p^{2}}\mathbf{p}\times\underline{\mathbf{S}}\right)  ,
\label{rP}%
\end{equation}
where $\alpha=\frac{1}{2}$ for fields and $-\frac{1}{2}$ for the vector
potential, \underline{$\mathbf{S}$} is the dimensionless spin-$1$ operator,
and $\underline{I}$ is the unit matrix. In our notation, underscore denotes a
matrix and bold denotes a $3$-component vector. \ Thus, the bold face
signifies that $\underline{\mathbf{r}}_{P}$ has $x$, $y$ and $z$ components,
while its underscore means that each of these components is a $3\times3$ array
that operates on the vector field of a first-quantized photon state, expressed
as a $3\times1$ array. This notation is carefully maintained to prevent
confusion between these two vector roles.

The Cartesian components of $\underline{\mathbf{r}}_{P}$ do not commute and
thus cannot define a basis of localized states. \ A family of position
operators that do have commuting components is%
\begin{equation}
\underline{\mathbf{r}}^{(\chi)}=i\hbar\underline{D}p^{\alpha}\nabla
p^{-\alpha}\underline{D}^{-1} \label{rD}%
\end{equation}
where $\underline{D}=e^{-i\underline{S_{3}}\phi}e^{-i\underline{S_{2}}\theta
}e^{-i\underline{S_{3}}\chi}$ is the rotation matrix with Euler angles
$\phi,\theta,\chi$ that rotates the lab $z$ axis into $\mathbf{\hat{p}.}$ The
role of the matrices \underline{$D$} and \underline{$D$}$^{-1}$ is to decouple
the spin and momentum, allowing the gradient operator to operate on the
momentum dependence of the field while maintaining the transversality
condition. A straightforward calculation gives \cite{HawtonBaylis}%
\begin{equation}
\underline{\mathbf{r}}^{(\chi)}=\underline{\mathbf{r}}_{P}-\hbar
\mathbf{a^{(\chi)}~}\widehat{\mathbf{p}}\cdot\underline{\mathbf{S}} \label{r}%
\end{equation}
with
\begin{equation}
\mathbf{a}^{(\chi)}\left(  \theta,\phi\right)  =\frac{\cos\theta}{p\sin\theta
}\widehat{\boldsymbol{\phi}}+\nabla\chi\left(  \theta,\phi\right)  . \label{a}%
\end{equation}
The polar and azimuthal angles are denoted $\theta$ and $\phi$ in momentum
space and $\vartheta$ and $\varphi$ in position space.\ 

As the basis vectors for the field and hence for the first-quantized photon
wave function, we use complex vectors \underline{$e$}$_{\lambda}$ of definite
helicity $\lambda=\pm1$, with components\
\begin{equation}
e_{\lambda,\mu}^{(\chi)}\left(  \theta,\phi\right)  =e_{\lambda,\mu}%
^{(0)}\left(  \theta,\phi\right)  \exp\left(  -i\lambda\chi\right)  .
\label{e_gamma}%
\end{equation}
where%
\begin{equation}
e_{\lambda,\mu}^{(0)}\left(  \theta,\phi\right)  =\left(  \widehat
{\boldsymbol{\theta}}_{\mu}+i\lambda\widehat{\boldsymbol{\phi}}_{\mu}\right)
/\sqrt{2} \label{e0}%
\end{equation}
and we add \underline{$e$}$_{0}=$\underline{$\widehat{\mathbf{p}}$} to
complete the triad. The ``hat''\ denotes a unit vector, and $\mu=-1,0$ and
$1,$ label rows of the column vector, \underline{$e$}$_{\lambda},$ and denote
components on the complex vectors $\left(  \underline{\widehat{x}}%
-i\underline{\widehat{y}}\right)  /\sqrt{2}$, $\underline{\widehat{z}}$ and
$\left(  \underline{\widehat{x}}+i\underline{\widehat{y}}\right)  /\sqrt{2},$
respectively,\ which are eigenvectors of \underline{$S$}$_{z}$ with eigenvalue
$\mu.$ Here we express the rotation matrix \underline{$D$} in terms of the
same components \cite{note1} and note that $e_{\lambda,\mu}^{\left(
\chi\right)  }\left(  \theta,\phi\right)  =D_{\mu\lambda}=D_{-\mu,-\lambda
}^{\ast}$. The general transverse basis vector \underline{$e$}$_{\lambda
}^{\left(  \chi\right)  }$ is rotated relative to \underline{$e$}$_{\lambda
}^{\left(  0\right)  }$ by the Euler angle $\chi$ about $\mathbf{p,}$ giving
just a phase difference in the helicity basis.

While the phase of the basis vectors depends on the choice of $\chi,$ the
physical fields are obviously independent of how we choose to orient the basis
vectors around $\mathbf{p}$. Indeed, we can rotate the basis vectors around
$\mathbf{p}$ by a different angle at different momentum-space positions
$\left(  \theta,\phi\right)  ,$ and this cannot change the physical field. In
this sense, a reorientation transformation $\chi\left(  \theta,\phi\right)
\rightarrow\chi^{\prime}\left(  \theta,\phi\right)  $ is a true local gauge
transformation. It is a basic requirement of the covariance of the geometric
representation. The invariance of the physical field and hence the photon wave
function means that the coefficients of the field when expanded in the basis
receive compensating phase factors \cite{note2}.

The term $\mathbf{a}^{\left(  \chi\right)  }$ may be considered an abelian
momentum-space vector potential, analogous to the vector potential
$\mathbf{A}$ of electromagnetic theory \cite{HawtonBaylis}. The position
operator \underline{$\mathbf{r}$}$^{\left(  \chi\right)  }$ in Eq.(\ref{r})
depends on the gauge of $\mathbf{a}^{\left(  \chi\right)  }$ through
$\nabla\chi\left(  \theta,\phi\right)  ,$ similar to the way the kinetic
momentum $\mathbf{P}$ of a massive charged particle depends on the gauge of
the vector potential $\mathbf{A.}$ The role of the charge of the massive
particle is seen to be taken in momentum space by the helicity of the photon,
and it is relevant to recall here the well-known result that the helicity
defines an invariant subspace of the Poincar\'{e} group. The basis vectors are
taken as eigenstates of the position operator, and a gauge transformation
cannot change their eigenvalues. Thus, a gauge transformation in the basis
states \underline{$e$}$_{\lambda}^{(\chi)}$ of the helicity subspace
$\lambda,$ say \underline{$e$}$_{\lambda}^{(\chi)}\rightarrow\underline
{e}_{\lambda}^{(\chi^{\prime})}=T\underline{e}_{\lambda}^{(\chi)}$ must change
the position operator according to the usual gauge rule%
\[
\mathbf{r}\underline{e}_{\lambda}^{(\chi)}\rightarrow\mathbf{r}^{\prime
}\underline{e}_{\lambda}^{(\chi^{\prime})}=T\mathbf{r}\underline{e}_{\lambda
}^{(\chi)}%
\]
and this gives the transformation $\mathbf{r}^{\prime}=T\mathbf{r}T^{-1}.$ In
our case, $T$ is the phase factor $T=e^{-i\lambda\left(  \mathbf{\chi}%
^{\prime}-\chi\right)  }$ and \underline{$\mathbf{r}$}$=i\hbar\underline
{D}p^{\alpha}\nabla p^{-\alpha}\underline{D}^{-1}$ so that $\mathbf{\underline
{r}}^{\prime}=\underline{\mathbf{r}}-\lambda\nabla\left(  \chi^{\prime}%
-\chi\right)  ,$ which is exactly the dependence we find for \underline
{$\mathbf{r}$} on the gauge transformation.

The field $\nabla\times\mathbf{a}^{\left(  \chi\right)  }$ in momentum space
corresponds to that of a magnetic monopole at the origin. The potential
$\mathbf{a}^{\left(  0\right)  }$ has singular ``Dirac''\ strings of flux
lines on the $\pm z$ axis that supply the flux emanating from the monopole.
This is most easily seen by integrating $\mathbf{a}^{\left(  0\right)  }$
along a path encircling the $z$ axis and equating this to the flux passing
through the area bounded by the path. The singular strings in $\mathbf{a}%
^{\left(  \chi\right)  }$ represent an essential nonintegrability or path
dependence that is responsible for the physical manifestation of the gauge
potential \cite{WuYang75}. As illustrated below, gauge transformations induced
by changes in $\chi\left(  \theta,\phi\right)  $ can change the strings, but
they do not alter the physical results. As shown in Ref. \cite{HawtonBaylis},
the abelian potential $\mathbf{a}^{\left(  \chi\right)  }$ is part of a more
general nonabelian gauge potential for SO(3).

The basis states (\ref{e_gamma}) can be used to express either the ideally
localized states (\ref{localstates}) or more readily realizable states.
Adlard, Pike and Sarkar \cite{APS}, for example, constructed single-photon
states with arbitrarily fast asymptotic power-law falloff of energy density
and photodetection rate and Bialynicki-Birula \cite{BBlocalization} obtained
converging or diverging localized states with an arbitrarily fast exponential
falloff. An advantage of these states is that the falloff rate for the vector
potential, the fields, and the Landau-Peierls photon wave function
\cite{LP}\ are asymptotically all determined by the same exponential factor,
and this avoids the problem that the fields themselves associated with exactly
localized states are not localized \cite{PikeSarkar}. As we show now, the
gauge choice $\chi\left(  \theta,\phi\right)  $ affects the angular momenta of
the basis states, whether\ applied to these asymptotically localized states or
to the exactly localized states of Ref. \cite{HawtonBaylis}.

The basis defined by
\begin{equation}
\chi\left(  \theta,\phi\right)  =-m\phi\label{chi_m}%
\end{equation}
has total $z$-angular momentum quantum number $j_{z}=m$ with the single-valued
gauge potential%
\begin{equation}
\mathbf{a}^{\left(  \chi\right)  }=\mathbf{\hat{\phi}}\frac{\cos\theta
-m}{p\sin\theta}. \label{a_chi}%
\end{equation}
The singularities in $\mathbf{a}^{\left(  0\right)  }$ along the $\pm z$ axis
($\theta=0,\pi)$ are thus changed in strength by the factors $1\mp m.$ For
example, for $m=1,$ the singularity along the positive $z$ axis is missing in
$\mathbf{a}^{\left(  \chi\right)  }$ whereas that along the negative $z$ axis
carries twice the flux. Other choices of $\chi\left(  \theta,\phi\right)  $
can reorient the singularity along some other direction or replace it by a
nonintegrable (multivalued) $\mathbf{a}^{\left(  \chi\right)  }.$\ \ A
reorientation of the singularity does not produce any new physics, and as
discussed above, for simplicity we choose a geometric gauge with the
singularity on the $\pm z$-axis. (The most general choice of $\chi$ can give a
singularity that is not straight as discussed in the literature on magnetic
monopoles \cite{GoddardOlive}, perhaps with interesting consequences.)
\ Restricting the Euler angle $\chi$ to functions given by Eq.(\ref{chi_m}),
the basis vectors can be expanded in eigenvectors of the usual spin-1 matrix
\underline{$S$}$_{z}$ and $L_{z}=-i\partial/\partial\phi$ as%
\begin{equation}
\underline{e}_{1}^{(-m\phi)}=\frac{1}{2}\left(
\begin{array}
[c]{c}%
\left(  \cos\theta-1\right)  \exp\left[  i\left(  m+1\right)  \phi\right] \\
-\sqrt{2}\sin\theta\exp\left(  im\phi\right) \\
\left(  \cos\theta+1\right)  \exp\left[  i\left(  m-1\right)  \phi\right]
\end{array}
\right)  \label{e_xyz}%
\end{equation}
The top row ($\mu=$ $-1$) gives the projection of the basis state
$\underline{e}_{1}^{(-m\phi)}$ onto a state with \underline{$S$}$_{z}%
$-eigenvalue $-1$ and $L_{z}$ eigenvalue $m+1$ with probability $\frac{1}%
{4}\left(  \cos\mathbf{\theta}-1\right)  ^{2};$ the second row ($\mu=0$), has
the corresponding eigenvalues $0$ and $m$ with probability $\frac{1}{2}\left(
\sin\mathbf{\theta}_{\mathbf{p}}\right)  ^{2}$, while the third row ($\mu=1$),
has eigenvalues $1$ and $m-1$ with probability $\frac{1}{4}\left(
\cos\mathbf{\theta}+1\right)  ^{2}$. Thus, by inspection, it is confirmed that
the total angular-momentum eigenvalue of \underline{$J$}$_{z}$ of the basis
state is $m$. \ The expectation values of $S_{z}$ and $L_{z}$ for the basis
state, obtained from the weighted sum, are then $\cos\theta$ and $-\cos
\theta+m$, respectively, showing that its cosine terms exactly cancel, leaving
the eigenvalue $m$ of \underline{$J$}$_{z}$ \ 

Restrictions on the uncertainty of the angular momentum of a localized state
are imposed by the commutation relations between the components of
$\underline{\mathbf{r}}^{(\chi)}$ and \underline{$\mathbf{J}$}, which were
found in Ref.\cite{HawtonBaylis} to be%
\begin{equation}
\left[  \underline{J}_{j},\underline{r}_{k}\right]  =i\epsilon_{jkl}%
\underline{r}_{k}-i\lambda\left(  \partial\underline{S}_{j}^{(\chi)}/\partial
p_{k}\right)  . \label{J_r_comm}%
\end{equation}
\ Note that the position operator does not transform as a simple vector
because, through its coupling to the spin, a rotation induces a gauge\ change.
\ For a photon at the origin for which $\left\langle \underline{r}%
_{k}\right\rangle =0,\ $the usual relationship between uncertainty and the
commutator gives%
\begin{equation}
\Delta\underline{J}_{j}\Delta\underline{r}_{k}\geq\frac{1}{2}\left\langle
\left|  \partial\underline{S}_{j}^{(\chi)}/\partial p_{k}\right|
\right\rangle \label{Jz-uncertainty}%
\end{equation}
and%
\begin{equation}
\Delta\underline{J}^{2}\Delta\underline{r}_{k}\geq\sum_{j}\left\langle \left|
\underline{J}_{j}\partial\underline{S}_{j}^{(\chi)}/\partial p_{k}\right|
\right\rangle . \label{J2-uncertainty}%
\end{equation}
When $\chi$ is given by Eq.(\ref{chi_m}) the $z$-component of $\underline
{\mathbf{S}}^{(\chi)}$ reduces to $\underline{S}_{z}^{(-m\phi)}=m\underline
{\mathbf{S}}\mathbf{\cdot\hat{p}}$ and within a state space of helicity
$\lambda,$ $\partial\underline{S}_{z}^{(-m\phi)}/\partial p_{k}=0.$ \ Thus the
photon can simultaneously have a definite position and $z$-component of the
total angular momentum. \ However, it does not have definite $x$ or
$y$-components of $\mathbf{J}$, and there is no definite value for the total
angular momentum. \ Nothing can be known definitely about the values of
\underline{$\mathbf{S}$} or \underline{$\mathbf{L}$} separately. \ This is
consistent with the expansion (\ref{e_xyz}). \ 

In coordinate space the electric field describing the localized states
discussed here can be written as%
\begin{align}
E_{\mu}\left(  \mathbf{r},t\right)   &  =\sum_{\lambda=\pm1}\int\frac{d^{3}%
p}{\left(  2\pi\hbar\right)  ^{3}}f\left(  p\right)  g\left(  \theta\right)
e_{\lambda,\mu}^{(\chi)}\left(  \theta,\phi\right) \nonumber\\
&  \times\exp\left[  i\left(  \mathbf{p\cdot r-}pct\right)  /\hbar\right]
\label{E}%
\end{align}
where $g\left(  \theta\right)  =\sin\theta$ for the localized states
considered in \cite{BBlocalization}, while $g\left(  \theta\right)  =1$ in
\cite{APS} and \cite{HawtonBaylis} and, with the gauge choice (\ref{chi_m}),%
\begin{equation}
e_{\lambda,\mu}^{(\chi)}\left(  \theta,\phi\right)  =e_{\lambda,\mu}^{(\chi
)}\left(  \theta,0\right)  e^{i\left(  m-\mu\right)  \lambda\phi}. \label{e}%
\end{equation}
To transform to coordinate space we use the expansion in spherical harmonics
\[
\exp\left(  i\mathbf{p}\cdot\mathbf{r}/\hbar\right)  =4\pi\sum_{l=0}^{\infty
}\sum_{n=-l}^{l}i^{l}Y_{l}^{n}\left(  \vartheta,\varphi\right)  Y_{l}^{n\ast
}\left(  \theta,\phi\right)  j_{l}\left(  pr/\hbar\right)
\]
and integrate over $\phi$ to obtain%
\begin{align}
E_{\mu}\left(  \mathbf{r},t\right)   &  =\frac{1}{\pi\hbar^{3}}\sum_{l=\left|
m-\mu\right|  }^{\infty}i^{l}Y_{l}^{m-\mu}\left(  \vartheta,\varphi\right)
\label{E-field}\\
&  \times\int d\left(  \cos\theta\right)  Y_{l}^{m-\mu\ast}\left(
\theta,0\right)  g\left(  \theta\right)  e_{\lambda,\mu}^{(\chi)}\left(
\theta,0\right) \\
&  \times\int dpp^{2}f\left(  p\right)  j_{l}\left(  \frac{pr}{\hbar}\right)
\exp\left(  -ipct/\hbar\right) \nonumber
\end{align}
where the subscript $\mu$ implies the corresponding component in the expansion
(\ref{e_xyz}).\ \ The position space field components vary as $\exp\left[
i\left(  m-\mu\right)  \varphi\right]  $, indicating a $z$-component of
orbital angular momentum equal to $\hbar l_{z}=\hbar\left(  m-\mu\right)  $.
\ Thus the position space $z$-components of spin, orbital and total angular
momentum are exactly the same as those in momentum space, and all of the
specific results discussed above regarding the angular momentum apply in
position space. \ 

The $\vartheta$ dependence can be obtained by expanding the integrand as
\begin{equation}
\sqrt{2\pi}g\left(  \theta\right)  e_{\lambda,\mu}^{(\chi)}\left(
\theta,0\right)  =\sum_{l=\left\vert m-\mu\right\vert }^{\infty}c_{\mu,l}%
Y_{l}^{m-\mu}\left(  \theta,0\right)  \label{theta-dependence}%
\end{equation}
and using the orthogonality of the spherical harmonics with the same
$\left\vert s_{z}\right\vert $ value. \ We consider a few examples. \ If $m=1$
and $g=1$ then $c_{0,l}=4/\sqrt{6}Y_{1}^{1}\delta_{l,1}$\ so that the
$z$-component of the field, $\sim\sin\vartheta.$ For the counterclockwise
rotating component of the field, $c_{1,l}=4/\sqrt{3}Y_{0}^{0}\delta_{l,0}$
$+2/\sqrt{6}Y_{1}^{0}\delta_{l,1}$ which gives a $\vartheta$- independent term
and a $\cos\vartheta$ term. \ The basis in Ref. \cite{BBlocalization} implies
$m=0$ and $g=\sin\theta$ and gives $c_{0,l}=4/\sqrt{3}Y_{0}^{0}\delta
_{l,0}-10/\sqrt{8}Y_{2}^{0}\delta_{l,2}.$ \ In all cases the field component
vanishes along any axis for which the corresponding component of $\mathbf{L}$
has a nonzero value.

In place of the common orbital angular momentum operator $\mathbf{L=}%
-\mathbf{p}\times i\nabla$ in momentum space, we should for consistency with
our position operator use%
\begin{equation}
\underline{\mathbf{L}}^{\left(  \chi\right)  }=\underline{\mathbf{r}}^{\left(
\chi\right)  }\times\mathbf{p}=\underline{D}\mathbf{L}\underline{D}%
^{-1}\mathbf{~.} \label{L-Chi}%
\end{equation}
The corresponding spin operator is%
\begin{equation}
\underline{\mathbf{S}}^{(\chi)}\equiv\mathbf{J-L}^{\left(  \chi\right)
}=\left(  \mathbf{a}^{(\chi)}\times\mathbf{p}+\widehat{\mathbf{p}}\right)
\underline{\mathbf{S}}\mathbf{\cdot}\underline{\mathbf{\hat{p}}} \label{S-Chi}%
\end{equation}
where $\underline{\mathbf{J}}\mathbf{=-}i\hbar\underline{I}\mathbf{p}%
\times\mathbf{\nabla\,+\,}\underline{\mathbf{S}}.$ The basis vectors are
eigenvectors of \underline{$\mathbf{S}$}$\cdot$\underline{$\mathbf{\hat{p}}$}
with eigenvalue $\lambda$, that is \underline{$\mathbf{S}$}$\cdot$%
\underline{$\mathbf{\hat{p}}$}$\underline{e}_{\lambda}=\lambda\underline
{e}_{\lambda}^{(\chi)}$. \ They are also eigenvectors of the position operator
with eigenvalue $0$, giving \underline{$\mathbf{r}$}$\mathbf{^{(\chi)}%
}\underline{e}_{\lambda}^{(\chi)}=0\ $and thus \underline{$\mathbf{L}$%
}$\mathbf{^{(\chi)}}\underline{e}_{\lambda}^{(\chi)}=0.$ Thus in a basis
expansion, \underline{$\mathbf{L}$}$^{(\chi)}$ just differentiates the
coefficient of \underline{$e$}$_{\lambda}^{(\chi)}$, giving no contribution
due to the basis. \ The operator $\underline{\mathbf{S}}^{(\chi)}$ alone
extracts the total angular momentum of the basis vector \underline{$e$%
}$_{\lambda}^{(\chi)}$. Thus use of the position operator, (\ref{r}) separates
the angular momentum of the basis from that in its coefficient.\ The
gauge-dependent term $\mathbf{a^{(\chi)}~}\widehat{\mathbf{p}}\cdot
\underline{\mathbf{S}}$ in the position operator (\ref{r}) can best be
understood in terms of its relationship to the angular momentum of the basis.
\ In position space, orbital angular momentum is associated with a component
of the Poynting vector in the $\widehat{\boldsymbol{\varphi}}$ direction such
that it spirals along the direction of propagation \cite{Allen92}. \ In
momentum space the functional forms of the position and momentum operators are
exchanged, and an analogous term appears in the position operator,
representing a spiraling of the field about $\mathbf{p.}$

The singular string of $\mathbf{a}^{\left(  \chi\right)  }$ discussed above is
the axis of a vortex. Expression (\ref{e_xyz}) makes explicit the angular
momenta of the basis vectors along the string, and associated clockwise and
counterclockwise rotation about it. The polar angle $\theta=0$ identifies the
positive $z$-axis and the paraxial limit when describing a beam, while
$\theta=\pi$ identifies the negative $z$-axis. \ If $m=0,$ the whole $z$-axis
is singular, while if $m=1,$ there is no singularity associated with the
positive $z$-axis ($l_{z}=0$), but the negative $z$-axis has $l_{z}=2,$ that
is it has twice the strength or topological charge. \ The singularity has just
been moved from the positive to the negative $z$-axis. \ The center of the
vortex has zero intensity due to the $\sin\theta$ dependence discussed above.
\ The orbital angular momentum arises from a bright annular ring about the
axis, as witnessed in the $j_{l}\left(  pr/\hbar\right)  \sin\vartheta$
dependence of the field (\ref{E-field}), and the radius of this ring goes to
zero with the parameter describing the spatial extent of the localized photon state.

In summary, each member of a family of position operators with commuting
components defines a corresponding basis of transverse unit vectors. \ The
choice of basis contributes a term $\lambda\mathbf{a}^{(\chi)}\times
\mathbf{p}$ to the total angular momentum of the basis states and affects the
associated optical vortex. \ However, a complete geometric gauge
transformation does not change the total field describing a physically
distinct photon state. \ The fact that the position operator given by
Eq.(\ref{r}) is not unique is a consequence of unavoidable ambiguity in the
selection of a transverse basis, implicit in the description of any exactly or
partially localized state.

\begin{acknowledgments}
The authors thank Juan Leon for private correspondence, Iwo Bialynicki-Birula
for stimulating discussions, and the Natural Sciences and Engineering Research
Council for financial support. \ 
\end{acknowledgments}


\begin{thebibliography}{9}
\bibitem {note1}In Ref. \cite{HawtonBaylis}, a Cartesian representation was
used for the matrices \underline{$S$}$_{j}$ and \underline{$D$}. The two
representations are related by a unitary transformation \underline{$T$},
Eq.(16) of Ref. \cite{HawtonBaylis}.
\bibitem {note2}It is implicitly assumed here that the oreintation of the
reference frame in the lab is independent of $\mathbf{\hat{p}}$. If it does
depend on $\mathbf{\hat{p},}$ as might be convenient in some scattering
experiments, then that part of the phase shift in the basis vectors is not
compensated by opposite shifts in the coefficients.
\end{thebibliography}
\end{document}